\documentclass[aps,prl,twocolumn,showpacs,groupedaddress]{revtex4}
\usepackage[dvips]{graphicx}
\usepackage{amssymb}
\usepackage{amsmath}
\usepackage{color}

\begin{document}
%\preprint{}

\title{Unveiling quantum Hall transport by Efros-Shklovskii to Mott variable range
hopping transition with Graphene}

\author{Keyan Bennaceur}

\author{Patrice Jacques}

\author{ Fabien Portier }

\author{P. Roche}

\author{D.C. Glattli}

\email{christian.glattli@cea.fr}

\altaffiliation{Also at Laboratoire Pierre Aigrain, 24 rue Lhomond, 75231 Paris Cedex 05, France}

\affiliation{Service de Physique de l'Etat Condens{é}/IRAMIS/DSM (CNRS URA 2464),
CEA Saclay, F-91191 Gif-sur-Yvette, France}

\date{\today}
\begin{abstract}
The quantum localization in the quantum Hall regime is revisited
using Graphene monolayers with accurate measurements of the
longitudinal resistivity as a function of temperature and current.
We experimentally show for the first time a cross-over from
Efros-Shklovskii Variable Range Hopping (VRH) conduction regime
with Coulomb interaction to a Mott VRH regime without
interactions. This occurs at Hall plateau transitions for
localization lengths larger than the interaction screening length
set by the nearby gate. Measurements of the scaling exponents of
the conductance peak widths with both temperature and current give
the first validation of the Polyakov-Shklovskii scenario that VRH
alone is sufficient to describe conductance in the Quantum Hall
regime and that the usual assumption of a metallic conduction
regime on conductance peaks is unnecessary.
\end{abstract}


\maketitle Since its recent discovery
\cite{Novoselov2005QHE,kimQhe}, the anomalous Quantum Hall Effect
(QHE) displayed by relativistic like electrons in a Graphene
monolayer has been mostly investigated to search for quantum Hall
ferromagnetism \cite{oddinteger} or Fractional Quantum Hall effect
\cite{AndreiFQHE,KimFQHE} in very low disorder samples to favor
interactions effects. Here we address the opposite regime where
disorder is strong enough to hide electron interactions, as it is
the case for standard exfoliated Graphene monolayers deposited on
the oxide layer of a silicon substrate. Graphene offers a new set
of parameters to revisit the quantum phase transition of
localization in the Quantum Hall regime. In particular, we show
that the screening of interactions by the $300\ $nm close
back-gate provided by highly doped silicon allows to observe for
the first time the transition from Efros-Shklovskii (E-S) to Mott
Variable Range Hopping (VRH) in QHE for large localization length.
The universal scaling exponents of quantum localization at Hall
plateau transitions deduced from our whole set of temperature and
bias current data, together with the E-S to Mott VRH transition,
definitely validate the Polyakov-Shklovskii (P-S) suggestion that
VRH transport is sufficient to describe the quantum Hall
electrical transport \cite{VRHshklovskii}.

The integer Quantum Hall effect occurs whenever Landau Levels
(LLs) which form due to quantization of carrier cyclotron orbits
are fully filled. In a conventional 2D-electron gas (2DEG)
positive energy LLs with $E_{n}=\hbar eB/m^{*}(n+1/2)$, $n$
integer, lead to LLs filling factors $\nu=k$, and quantized Hall
resistance $R_{H}=h/ke^{2}$ with $k$ a positive integer and spin
degeneracy lifted assumed. In a Graphene monolayer, carriers obey
a ultra-relativistic Dirac equation. LLs have both positive and
negative energies $E_{n}=\pm hV_{F}/l_{c}\sqrt{(2n)}$ where
$V_{F}$ is the Fermi velocity and $l_{c}=\sqrt{(\hbar/eB)}$ the
magnetic length. Because of the existence of a zero kinetic energy
LL and the non lifted four fold valley and spin degeneracy, the
filling factors series leading to Hall plateaus becomes
$\nu=\pm4(k+1/2)$. The finite width of the quantized Hall plateaus
occurs because of localized states in the bulk provided by
potential disorder whose energies spread around the unperturbed LL
energies. For filling factors yielding the Hall plateau series
mentioned above, the Fermi level $E_{F}$ lies between two
unperturbed LLs, i.e. on localized state energies. The
longitudinal conductance $\sigma_{xx}$ vanishes at zero
temperature, preventing electron backscattering through the bulk
and ensuring perfect quantization of the Hall current circulating
on the sample edges. The transition between two Hall plateaus
occurs when the Fermi level lies in the middle of a disorder
energy broadened LL. For an infinite size sample, the localized
state size $\xi$ diverges at a single energy $E_{c}\approx E_{n}$
resulting in backscattering and a longitudinal conductance peak.
According to the quantum localization theory
\cite{ReviewPercolation}
$\xi\sim|E-E_{c}|^{-\gamma}\sim|\nu-\nu_{c}|^{-\gamma}$ with
$\gamma\simeq7/3$, a value confirmed in many experiments on
conventional 2DEGs \cite{PruiskenWei,furlan}.

Deducing the scaling exponent from transport measurement requires
a model linking transport quantities to $\xi$. In general one can
write $\sigma_{xx}=f(\xi/L(T))$ where $f$ is a universal scaling
function and $L(T)$ a characteristic length depending on the
conduction mechanisms at finite temperature. Let us first discuss
bulk conduction on the Hall plateaus. It is generally accepted
that transport occurs via phonon assisted inelastic transitions
between localized states, the so-called variable range hopping
mechanism. For non-interacting electrons, the VRH  Mott's law
gives
\begin{equation}
\sigma_{xx}\propto\frac{T_{M}}{T}\exp(-(T_{M}/T)^{1/3})\label{eq:MottVRHT}\end{equation}
or equivalently
$\sigma_{xx}\propto(\frac{\xi}{L_{M}(T)})^{2}\exp(-(L_{M}(T)/\xi)^{2/3})$,
which defines the characteristic length $L(T)$ labeled as
$L_{M}(T)=\sqrt{1/\pi g(\varepsilon_{F})k_{B}T}$ , where
$g(\varepsilon)$ is the energy independent density of states at
the Fermi energy. However, in the QHE regime screening is poor and
Coulomb repulsion must be included. One thus enters the
Efros-Shklovskii (E-S) VRH regime, where the density of states
$g(E)\propto|E-E_{F}|$ yields
\begin{equation}
\sigma_{xx}\propto\frac{T_{0}}{T}\exp(-(T_{0}/T)^{1/2})\label{eq:ESVRHT}\end{equation}
or
$\sigma_{xx}\propto(\frac{\xi}{L_{E-S}})\exp(-(L_{E-S}/\xi)^{1/2})$\cite{efrosshklovskii}
, with both the length $L_{E-S}(T)=4\pi\varepsilon_{0}\varepsilon
k_{B}T/Ce^{2}$ and the energy
$k_{B}T_{0}=Ce^{2}/4\pi\varepsilon\varepsilon_{0}\xi$ given by the
Coulomb energy. $C\simeq6.2$ is a numerical constant
\cite{CksiShklovskii}. Measuring $T_{0}$ thus allows to determine
$\xi$ and to probe the scaling law far from the conductance peaks.
Still in the same regime, passing a current $I$ trough the sample
while keeping a fixed low temperature gives an E-S VRH like law
for $\sigma_{xx}$ where the current plays the role of the
temperature. This is the P-S model \cite{PolyakovSchklovskii}
which uses the effective electronic temperature $k_{B}T\rightarrow
eE_{H}\xi/2$ where the local Hall electric field $E_{H}$ is
proportional to the current $I$. This leads to \begin{equation}
\sigma_{xx}\propto
\exp(-((E_{0}/E_{H})^{1/2})\label{eq:ESVRHI}\end{equation}
 and \begin{equation}
\sigma_{xx}\propto
\exp(-((E_{1}/E_{H})^{1/3})\label{eq:MottVRHI}\end{equation}
 for E-S and Mott's VRH respectively.

Probing the localization scaling law for large localization length
requires understanding the conduction mechanism close to the
conductance peaks, in the plateau transition region where the
Fermi level approaches a unperturbed LL. Historically, the first
conduction mechanism proposed for conductance peaks was a metallic
regime. The Pruisken model \cite{Pruisken} sets the characteristic
length $L(T)$ as the phase coherence length
$L_{\phi}(T)=(D\tau_{\phi})^{1/2}$. Here $D$ is a diffusion
constant and the phase coherence time $\tau_{\phi}\propto T^{-p}$
follows a non-universal power law with $T$ ($p=2$ accounts for
most observations). The localization scaling exponent $\gamma$ can
be indirectly accessed by the temperature dependence of the Full
Width at Half Maximum (FWHM) $\Delta\nu$ of the conductance peaks.
The latter is obtained when $\xi(\Delta\nu/2)\simeq L_{\phi}(T)$
giving $\Delta\nu=(T/T_{1})^{\kappa}$ with the non universal
exponent $\kappa=p/2\gamma$. However Polyakov and Shklovskii
proposed that the VRH regime should last in the plateau transition
region and the FWHM obtained from $\xi(\Delta\nu/2)\simeq
L_{E-S}(T)$ (or $T\simeq T_{0})$ ) giving
$\Delta\nu=(T/T_{1})^{\kappa}$ and the now universal
$\kappa=1/\gamma$. Here
$k_{B}T_{1}=Ae^{2}/4\pi\varepsilon\varepsilon_{0}\xi$ with $A$ a
numerical constant. Similarly, the dependence of the FWHM with
bias current using P-S model is $\Delta\nu=(I/I_{1})^{\mu}$ with
$\mu=1/2\gamma=\kappa/2$ while using the phase coherence length
approach $\mu=p/4\gamma$. From the latter discussion we see that,
as $p=2$ is a reasonable exponent for $\tau_{\phi}$, the scaling
law of the FWHM with temperature can not discriminate between the
two scenariios nor that with bias current.

A further criterion is thus needed to discriminate the two
scenarios, and that is addressed experimentally in this Letter.
The idea originates from the Aleiner Shklovskii (A-S)
\cite{SchklovskiiScreening} prediction that a cross-over from E-S
to Mott VRH occurs when interactions are screened, for example by
a gate parallel to the 2DEG at a distance $d$. This requires
$\xi>2d$ which is likely to occur on conductance peaks for sample
size $\gg2d$. In the Mott VRH regime the FWHM conductance peak now
becomes $\Delta\nu\propto(T/T_{2})^{\kappa}$ with
$\kappa=1/2\gamma$ and $\Delta\nu\propto(I/I_{1})^{\mu}$ with
$\mu=1/3\gamma$. On the contrary in the Pruisken scenario
screening is not expected to impact the temperature dependence of
the FWHM. This yet never observed E-S to Mott cross-over in the QH
regime would definitely establish the P-S scenario, that VRH
describes transport almost everywhere in the QHE regime even close
to the maximum of the conductance peaks  and that the phase
coherence length approach is not appropriate.

Previous measurements performed in conventional 2DEGs, including
Si-MOSFETs, and InAs/InGAAs or GaAs/AlGAAs heterojunctions have
been able to probe the scaling exponent of $\xi$. Experiments
using direct determination of $\xi$ from the E-S VRH \cite{furlan}
and even more directly by geometrical comparison with sample width
\cite{KlitzingScalingSize} have given $\gamma\simeq2.3$. Probing
the scaling law using the conductance peak width is less direct
and showed a dispersion in the extracted values of $\kappa$. Works
combining temperature and bias current have shown excellent
agreement with the P-S model
\cite{Hohls,furlan,WeiCurrent,KlitzingScalingSize}. Recently the
scaling law has been studied in Graphene using temperature, but no
current bias study was done \cite{giesbersscaling}. Except on the
$n=0$ LL level the results were found compatible with
$\gamma=2.3$. So far no experiments in the QHE have shown the E-S
to Mott cross-over. It has been only observed in highly disordered
2D electron systems in zero field \cite{Peper2DVRH}. Here, the
cross-over occurs not because of screening but for energies well
above the Coulomb gap, restoring a constant density of states.

In this paper we present a complete set of data in temperature and
bias current on the QHE regime performed on Graphene monolayers.
The silicon back-gate with $d=300$nm gives for the first time
access to the cross-over from Mott to ES VRH regime on
conductivity peaks for the highest filling factors where $\xi$ is
large. All the scaling exponents $\gamma,\kappa$ and $\mu$ agree
with the P-S and A-S prediction in the screened and unscreened
regime. This provides a definitive confirmation of these models.
The scaling exponent found to be $\gamma\simeq2.3$ is the same for
first two $n=\pm1,\pm2$ LLs including the $n=0$ LL where no
anomalous behavior is observed as found in
Ref.\cite{giesbersscaling}.

Four samples (S1 to S4) have been fabricated using exfoliation of
natural graphite flakes \cite{novoselov-2004-306} . All samples
have been deposited on the 300nm thick oxide layer of highly doped
silicon wafer which serves as a back gate. Contacts where made
using e-beam lithography and evaporating 5/70 nm Ti/Au in high
vacuum. After processing, S1 and S2 were covered with PMMA while
S3 and S4 were not. Samples S3 and S4 were heated up to 450K in
cryogenic vacuum during several hours until the four points
resistance reached a steady high value signaling low dopant
concentration. Mobilities at $1.10^{12}$cm$^{-2}$ are around
$3000$ cm$^{2}V^{-1}s^{-1}$ for S1 and S2, $6000$
cm$^{2}$V$^{-1}$s$^{-1}$ for S3 and $10000$
cm$^{2}$V$^{-1}$s$^{-1}$ for S4. Sample S4 showed the quantum Hall
plateaus series and a Raman spectrum of a monolayer although the
gate efficiency was two times smaller than expected. It is likely
a twisted bilayer.

The longitudinal resistivity $\rho_{xx}$ was recorded while
varying density for about 50 temperature values ranging from 1.6K
(S4) or 4K (S3) to 300K and similar runs are repeated for
different fixed high magnetic fields from 6 to 17 Tesla. These
extensive measurements have been done on two samples (S3 and S4).
Measurements as a function of bias current were also performed
ranging from $10nA$ to $100\mu A$ at fixed temperature (4.2K or
1.6K). These measurements have been done on the four samples. Four
point measurements were done on S3 $(R_{xx}$) and S4 $(R_{xx}$ and
$R_{hall})$ whereas 3 point measurements were performed on the
other two samples. In order to extract the resistivity from the
resistance the aspect ratio was estimated using a numerical
electrostatic calculation. For all measurements the density is
slowly swept (less than 2 V/mn ) to avoid hysteretic effects due
to trapped charges in silicon oxide. To get rid of the weak
remaining hysteresis the data are taken always in the same
sweeping direction. Most of the data presented in this Letter are
from sample S3 and some of the data from sample S4. Other samples
were used to confirm the reproducibility of the physical
properties.

\begin{figure}[h]
\centerline{\includegraphics[width=8.5cm]{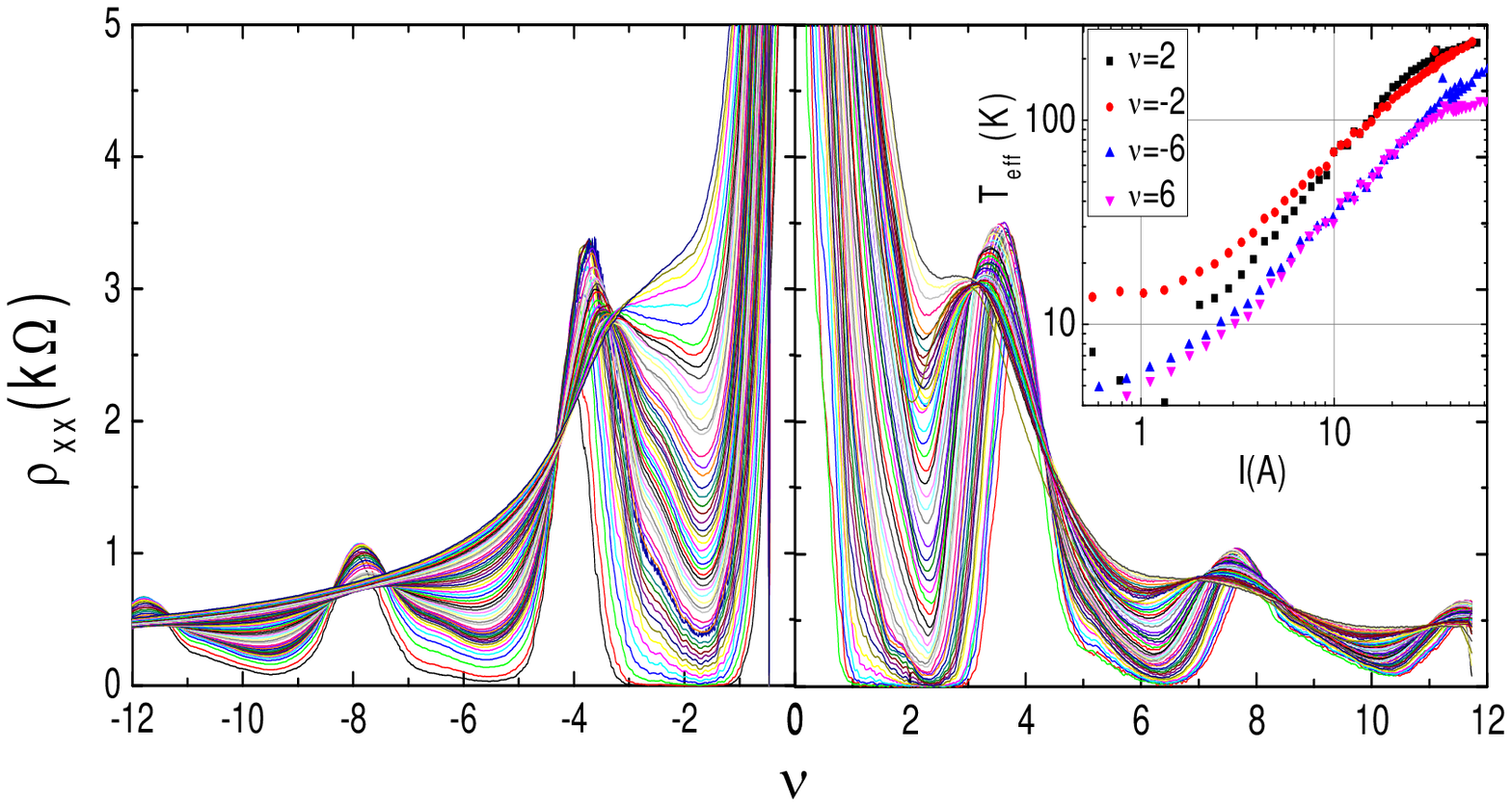}}

\caption{Longitudinal resistance $\rho_{xx}$ versus filling factor
at 16.5T. On the left the temperature is varied from 4.2K to 300K
and on the right the bias current from $10$nA to
$100\mathrm{\mu}$A at 4.2K. Inset: effective temperature
$T_{eff}(I)$ (see definition in text) versus bias current $I$ for
different filling factors on plateaus. } \label{figure1}
\end{figure}

\begin{figure}[h]
\centerline{\includegraphics[width=7cm]{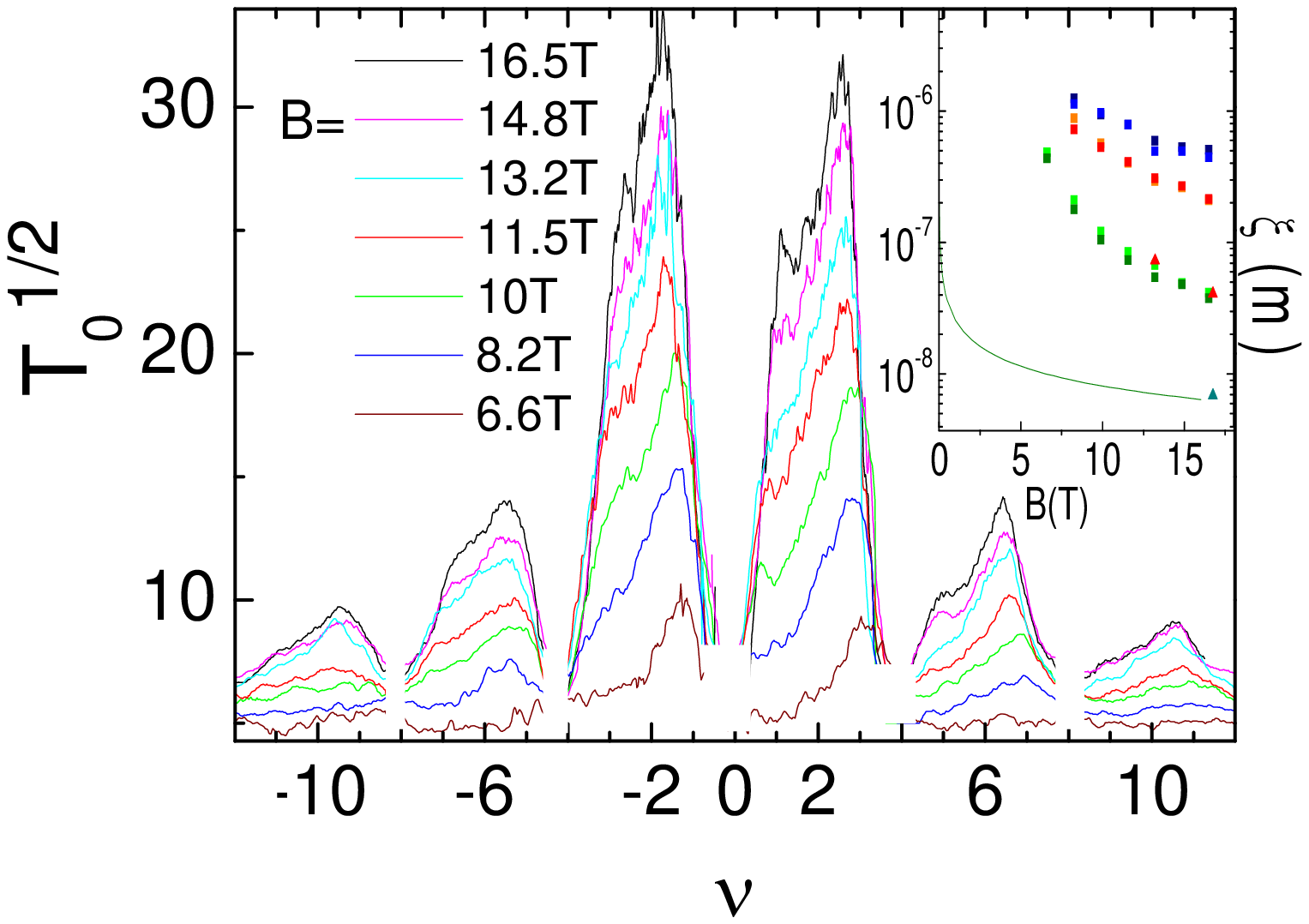}}

\caption{Square root of E-S VRH temperature $T_{0}$ versus filling
factor for sample S3 at different magnetic fields ranging from
6.6T to 16.5T. The inset shows $\xi_{min}$ as a function of the
magnetic field. Green, red, and blue data are for
$\nu=\pm2$,$\pm6$, and $\nu=\pm10$ respectively. Squares are for
sample S3 and triangles for S4. The solid curve corresponds to
$l_{c}$.}

\label{figure2}
\end{figure}

We now present the experimental results. An example of $\rho_{xx}$
measurements is shown in Figure \ref{figure1} at 16.5 Tesla for
S3. The left part shows gate voltage sweeps at low bias for about
50 different fixed temperatures and the right part for a series of
fixed current bias at base temperature. The whole data for both
positive and negative gate voltages can be found in the supporting
material \cite{SM}.
Let us first focus on the Hall plateau regions. The variation of
$\sigma_{xx}$ with $T$ was shown to follow accurately the E-S VRH
law \cite{VRHshklovskii}, see supporting material \cite{SM}. This
is well obeyed from 1.6K to $\simeq$ 80K throughout the $\nu=\pm2$
plateaus for all magnetic fields. We emphasize that both the
simple activated law and the 2D Mott's VRH law yields poor fits.
Above 100K departure from the E-S VRH law signals thermal
activation to the next LLs. From the fit combining activated and
E-S VRH law we can extract both the VRH temperature $T_{0}$ as
shown in Fig.\ref{figure2} and the activation energy $\Delta$. The
observed $\Delta$ are smaller than those given by the Dirac
equation in magnetic field
$\Delta=E_{n+1}-E_{n}=\frac{1}{2}(\sqrt{n+1}-\sqrt{n})\sqrt{2e\hbar
v_{F}^{2}B}$ (solid lines) because of LL disorder broadening. At
17T, 200K and 70K LL broadening are found respectively for sample
S3 and S4. The values for S3 are comparable with those obtained in
\cite{giesbersActivation}, see \cite{SM} for a complete set of
data.

The dependence with current bias shows accurate agreement with the
E-S VRH like law, Eq.\ref{eq:ESVRHI}, following the P-S
prediction. Comparing Eqs. \ref{eq:ESVRHT} and \ref{eq:ESVRHI} we
can write
$(\frac{\varepsilon_{H0}}{\varepsilon_{H}})^{\alpha_{1}}=(\frac{T_{0}}{T})^{\alpha_{2}}+k$,
where k is a constant. If the E-S law is obeyed for bias currents
one should have $\alpha_{1}=\alpha_{2}=1/2$, where
$\alpha_{2}=1/2$ was already established. In order to check this,
we define the effective temperature $T_{eff}(I)$ such that
$\rho_{xx}(I)=\rho_{xx}(T_{eff})$. On Fig.\ref{figure1}, right
inset, $T_{eff}$ is plotted as a function of the bias current in
the logarithmic scale for $\nu=\pm2,\pm6$. It is clear that below
100K (no thermal activation) $T_{eff}\propto I$ showing that
$\alpha_{1}=\alpha_{2}$ and a VRH like law for current is well
obeyed by $\sigma_{xx}$.
Fig.2 shows $T_{0}^{1/2}$ extracted from fits of $\sigma_{xx}(T)$
using Eq.3 for a continuous series of filling factors except on
the very maximum of the conductance peaks. From these measurements
we can extract the localization length via $T_{0}$ using:
$\xi(\nu)=Ce^{2}/4\pi\epsilon_{r}\epsilon_{0}k_{B}T_{0}(\nu)$. In
the inset of Fig.\ref{figure2} the smallest localization length
$\xi_{min}$ found in the middle of the plateaus is plotted as a
function of the magnetic field. The solid lines show the magnetic
length $l_{c}$ for comparison. In sample S3, $\xi_{min}\sim 40$ nm
for $\nu=\pm2$ at $B=16.5$ Tesla is around seven times larger than
$l_{c}$ whereas in sample S4 $\xi_{min}$ is of the order of
$l_{c}$. Smaller localized states can be expected due to higher
mobility of S4 or, if it is a twisted bilayer, due to screening of
the silicon oxide charge impurities by the lower layer. $\xi$
values in S3 are consistent with measurements of Ref.\cite{yacoby}
but quite below those found in
Refs.\cite{giesbersscaling,noteC=1}.

\begin{figure}[h]
 \centerline{\includegraphics[width=7cm]{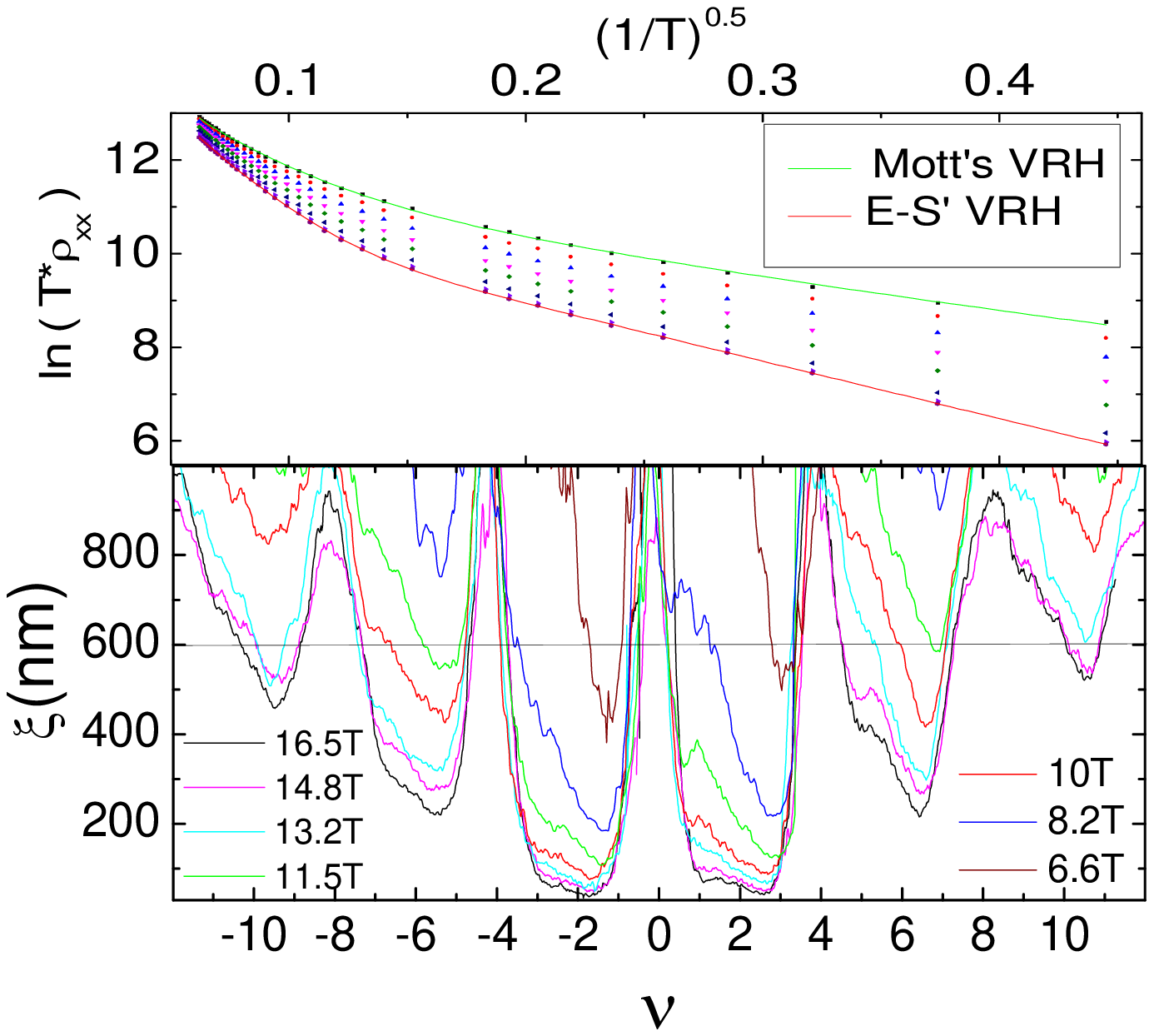}}

\caption{Lower graph: localization length $\xi$ versus filling
factor for S3 for different magnetic fields. The horizontal solid
line $\xi=600$nm shows the threshold above which screening of
interactions by the gate plays a role. Upper figure: plots of
$T\rho_{xx}$ versus $1/\sqrt{T}$ at 14.8T for various filling
factors near $\nu=-6$ and $\xi$ above and below $600$nm. The red
solid curve is a E-S law's fit of the data while the green solid
curve uses Mott's law. }

\label{figure3}
\end{figure}

In Fig.\ref{figure3}, lower graph, we show $\xi(\nu)$ extracted
from a E-S VRH analysis of the data for continuous values of the
filling factor and for different magnetic fields. The line at
$\xi=2d=600$nm signals the limit of validity for $\xi$ extracted
from the E-S VRH law. Indeed for larger $\xi$ we expect screening
of the interactions and a cross over from E-S to Mott VRH law.
This is what is observed as shown by a log plot of $T\sigma_{xx}$
as a function of $1/\sqrt{T}$ on the upper part of \ref{figure3}.
For $\xi<600$nm a linear variation is found while for $\xi>600$nm
the variation is no longer linear and is well fitted by the Mott's
law. This yet never observed cross-over from E-S to Mott's law is
one of the two main results of our experiment.

\begin{figure}[h]
 \centerline{\includegraphics[width=7cm]{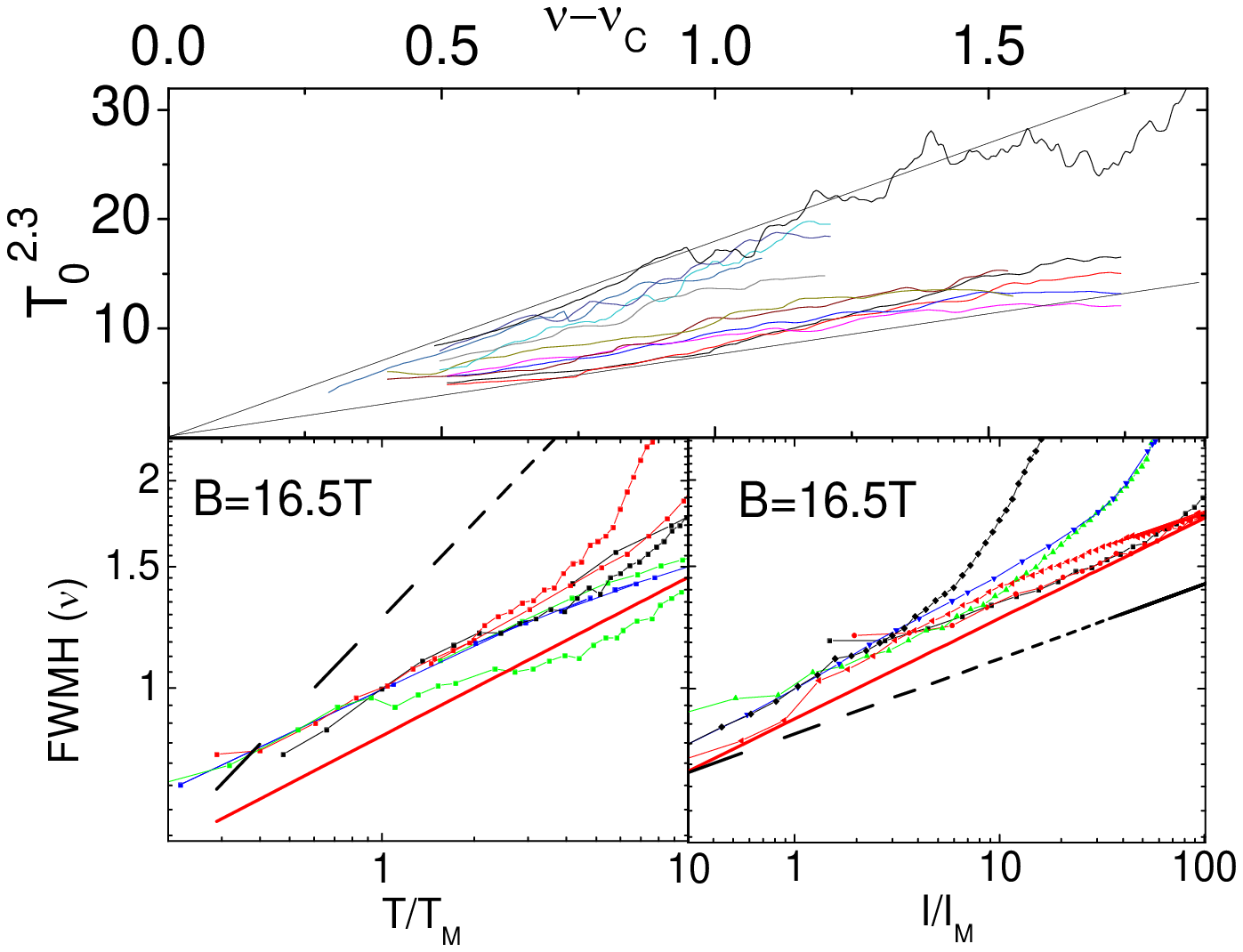}}

\caption{Upper figure: $T_{0}^{-2.3}$ as a function of $\nu-\nu_{C}$
for $\nu=[\pm2,\pm6]$ and $\nu=[\pm2,0]$ at different magnetic field,
the black curve is for S4 while the others are for S3. Lower part:
FWHM of the $\rho_{xx}$ peaks for temperature measurement
(left) and for bias current measurement (right). Blue, green and red curves
are data for the peaks between $\nu=\pm10$ and $\pm6$, $\nu=\pm6$ and $\pm2$, and $\nu=-2$ and $2$ respectively. Squares are for S3 and triangle for S4. The solid red and
dashed black lines correspond to the expected exponent for
E-S scenario and for the A-S scenario in the Mott regime respectively. The Pruisken scenario is also represented by the dashed black lines.}

\label{figure4}
\end{figure}

Before going further in the study of the Mott's regime, it is important to
determine the universal scaling exponent $\gamma$ in the E-S VRH regime
for which $\xi$ can be reliably known.
Fig. \ref{figure4}, upper graph, shows $T_{0}^{2.3}$ as a function
of $\nu-\nu_{C}$ for $\nu_{C}\simeq\pm2$ for samples S3 and S4.
The linear variation indicates that $\gamma=2.3$ is a reasonable
exponent. This result is in agreement with Ref. \cite{giesbersscaling}.

The second important result of this work comes from the study of
the scaling exponent of the FWHM $\Delta\nu$ of resistivity peaks
between Hall plateaus with both temperature and bias current.
Here, as $\xi>2d$ in this regime, Mott's VRH law is obeyed and we
expect the exponent values will allow to discriminate between the
E-S VRH and the phase coherence length scenario. The FWHM of resistance peaks
are plotted for both S3 and S4 on Fig.\ref{figure4} on the lower
left and lower right part respectively for
$\nu=[\pm2,\pm6],[\pm10,\pm6],[-2,+2]$. The figures clearly show
a universal behavior of $\Delta\nu$ at temperature below
100K and bias current below $10\mu\mathrm{A}$. $\kappa$ is found
to be equal to $0.23\pm0.02$ and $\mu$ to $0.13\pm0.01$ which are
both in good agreement with Mott's VRH confirming Mott's law at
the edge of the quantum Hall plateaus. If the phase coherence
length approach was relevant, we would have found $\kappa=0.42$
and
$\mu=0.21$. %

To conclude we have studied the quantum localization in the
quantum Hall regime using Graphene monolayers. The standard
localization length scaling exponents was found for all studied
Landau Levels. More important, our first observation of an
Efros-Shklovskii to Mott VRH cross-over in the quantum Hall regime
found on conductance peaks allows to discriminate between the
Polyakov-Shklovskii and the Pruisken phase coherence length
scenarios describing the conduction on the plateau transition.

We thank J-N Fuchs and M.Goerbig for fruitful discussions as well
as K. Novoselov for help in the early stage of the project. We
thank the Quantronics group and especially D. Vion for help in
nanofabrication. The CNano IdF grant JoseqhGraphn is acknowledged.


\begin{thebibliography}{10}

\bibitem{Novoselov2005QHE}
K.~S. Novoselov et al.,
\newblock {Nature}, \textbf{438}, 197, (2005).

\bibitem{kimQhe}
Y~Zhang, J~W Tan, H~L Stormer, and P~Kim,
\newblock {Nature}, \textbf{438}, (2005).

\bibitem{oddinteger}
Y.~Zhang et al.,
\newblock {Phys. Rev. Lett.}, \textbf{96}, 136806 (2006).

\bibitem{AndreiFQHE}
Xu~Du et al.,
\newblock {Nature}, \textbf{462}, 10 (2009).

\bibitem{KimFQHE}
K.~I. Bolotin et al.,
\newblock { Nature}, \textbf{462}, 196 (2009).

\bibitem{Peper2DVRH}
I.Shlimak and M. Pepper,
\newblock {Philosophical Magazine Part B}, \textbf{81}, 1093 (2001) and references therein.

\bibitem{VRHshklovskii}
D.~G. Polyakov and B.~I. Shklovskii,
\newblock { Phys. Rev. Lett.}, \textbf{70}, 3796 (1993).

\bibitem{ReviewPercolation}
see the review by B. H\"{u}ckenstein,
\newblock { Rev. Mod. Phys}, \textbf{67}, 357 (1995), and references therein.

\bibitem{PercolationLuryi}
R. F. Kazarinov and S. Luryi,
\newblock { Phys. Rev. Lett.}, \textbf{25}, 7626 (1982).

\bibitem{PruiskenWei}
H.~P. Wei, D.~C. Tsui, and A.~M.~M. Pruisken,
\newblock {Phys. Rev. B}, \textbf{33}, 1488 (1986).

\bibitem{KlitzingScalingSize}
S. Koch, R. J. Haug, K. V. Klitzing, K. Ploog,
\newblock {Phys. Rev. B}, \textbf{46}, 1596 (1992).

\bibitem{KlitzingScaling}
S. Koch, R. J. Haug, K. V. Klitzing, K. Ploog,
\newblock {Phys. Rev. Lett}, \textbf{67}, 883 (1991).

\bibitem{furlan}
M. Furlan,
\newblock {Phys. Rev. B}, \textbf{57}, 14818 (1998).

\bibitem{efrosshklovskii}
A.L Efros, B.I Shklovskii
\newblock {J. Phys. C}, \textbf{8}, 249 (1975).

\bibitem{CksiShklovskii}
B.I. Shklovskii,  A.L. Efros, and N. Van~Lien,
\newblock {Solid State Communications}, \textbf{32}, 851 (1979).

\bibitem{PolyakovSchklovskii}
D.~G. Polyakov and B.~I. Shklovskii,
\newblock {Phys. Rev. Lett.}, \textbf{70}, 3796 (1993).

\bibitem{Pruisken}
A.~M.~M. Pruisken,
\newblock {Phys. Rev. Lett.}, \textbf{61}, 1297 (1988).

\bibitem{SchklovskiiScreening}
I.~L. Aleiner and B.~I. Shklovskii,
\newblock {Phys. Rev. B}, \textbf{49}, 13721 (1994).

\bibitem{WeiCurrent}
H.~P. Wei, L.~W Engel, and D.~C. Tsui,
\newblock {Phys. Rev. B}, \textbf{50}, 14609 (1994).

\bibitem{giesbersscaling}
A.~J.~M. Giesbers et al.,
  \newblock {Phys. Rev. B}, \textbf{80}, 241411 (2009).

\bibitem{Hohls}
F.~Hohls, U.~Zeitler, and R.~J. Haug,
\newblock {Phys. Rev. Lett.}, \textbf{88}, 036802 (2002).

\bibitem{novoselov-2004-306}
K.~S. Novoselov et al.,
\newblock {Science}, \textbf{306}, 666 (2004).

\bibitem{SM}
see supporting material.

\bibitem{giesbersActivation}
A.~J.~M. Giesbers et al.,
\newblock {Phys. Rev. Lett.}, \textbf{99}, 206803 (2007).

\bibitem{yacoby}
J. Martin et al.,
\newblock {Nature Physics}, \textbf{5}, 669 (2009).

\bibitem{noteC=1}
in reference \cite{giesbersscaling} $C=1$ is used.

\end{thebibliography}
\end{document}